\documentclass[12pt]{article}

\usepackage{scicite}

\usepackage{times}
\usepackage{graphicx}
\usepackage{graphics}
\usepackage{pdflscape}
\usepackage{lscape}
\usepackage{graphicx}

\title{Virtual water controlled demographic growth of nations}

\author{ Samir Suweis$^{1}$, Andrea Rinaldo$^{2,3}$,
Amos Maritan$^1$, Paolo D'Odorico$^{4,5}$
\\
\normalsize{$^1$ Department of Physics, University of Padua, via Marzolo 9, 35131 Padova (Italy),}\\
\normalsize{$^2$Laboratory of Ecohydrology, EPFL, 1015 Lausanne,
(Switzerland)}\\
\normalsize{$^3$Dipartimento ICEA, University of Padua, Padova, 35131, Italy,} \\
\normalsize{$^{4}$ Department of Environmental Sciences, University of Virginia, VA 22904 (USA)}\\
\normalsize{$^{5}$EFLUM Laboratory, EPFL, 1015 Lausanne,
(Switzerland)}}
\begin{document}
\date{}
\maketitle


\begin{abstract} 
\textbf{Population growth is in general constrained by food production,
which in turn depends on the access to water resources. At a
country level, some populations use more water than they control
because of their ability to import food and the virtual water
required for its production. Here, we investigate the dependence
of  demographic growth on available water resources for exporting
and importing nations. By quantifying the carrying capacity of
nations based on calculations of the virtual water available
through the food trade network, we point to the existence of a
global water unbalance. We suggest that
current export rates will not be maintained and consequently we
question the long-run sustainability of the food trade
system as a whole. Water rich regions are likely to soon reduce
the amount of virtual water they export, thus leaving
import-dependent regions without enough water to sustain their
populations. We also investigate the potential impact of
possible scenarios that might mitigate these effects through (1)
cooperative interactions among nations whereby water rich
countries maintain a tiny fraction of their food production
available for export; (2) changes in consumption patterns; and (3)
a positive feedback between demographic growth and technological
innovations. We find that these strategies may indeed reduce the
vulnerability of water-controlled societies.}
\end{abstract}



\section{Introduction}
Most of the water we use is to produce the food we eat. With the
world's population  that has doubled every 40 years there is a
growing concern that water limitations  will soon impede
humanity to meet its food requirements (1-7). The urgency
to deal with this  alarming situation by developing durable
socio-political and economic strategies that promote a sustainable
use of the environment and its natural resources, was at the core
of the recent Rio+20 Earth Summit organized by United Nations (7-8). In recent
years a number of studies have combined projections of population
growth with predictions of water availability and agricultural
productivity under a variety of climate change and land use
scenarios. These predictions have been used to assess whether
mankind will run out of water in the next few decades, and to
investigate possible strategies to deal with the global water and
food crisis (9-15). At the regional scale, there exist several
areas of the world where the demand has already exceeded the
supply of renewable freshwater resources (16-19). How can this
negative water budget be sustained? Mainly by importing food. The
import of food commodities is associated with a virtual transfer
of freshwater resources from production to consumption areas (20).
Virtual water trade allows some populations to exceed the limits
imposed by their local water resources (14, 21-22). By sustaining
demographic growth above the regional carrying capacity,  virtual
water trade has mitigated the effects of drought and famine in
many regions of world (14,23). Thus, the redistribution of virtual
water (VW) resources  often appears as a remedy to regional water
crises (22-23). Even though, presently, most exporting countries
can afford to sustain VW exports, their demographic growth might
soon limit the amount of VW resources they can place on the global
market. At some point these societies will have to decide whether
they  want to sustain the existing export rates, or prefer to
reduce the exports in order to meet their own food demand. What
are the global implications of these two scenarios? To answer to 
this question, we need to relate demographic growth both to the available freshwater resources and to global patterns of VW trade. We first investigate current trends of demographic growth and then develop model-based predictions of how population is expected to change as water rich countries start reducing their exports. 

\section{The Carrying Capacities of Nations}
We classify all countries around the world into  five groups,
depending on their supply and demand of VW, and on the resulting
balance or unbalance between available and consumed water
resources (see Figure 1).  We observe that the water-rich regions
are in North and South America, Australia, and the former Soviet
Union (or "eastern block"). These regions are known for being
major VW exporters (26,27). Virtual water dependent regions (i.e.,
regions that need VW imports to meet their demand) are mainly in
Europe, Mexico, and the western side of South America. Despite VW
trade, large parts of Africa and Asia remain affected by water
stress. Because food production is the major form of freshwater
consumption by human societies (24-25), we calculate the "carrying
capacity" of a nation, i.e. its maximum sustainable population,
based on the water resources currently available for agriculture
and livestock. The carrying capacity, though difficult to
quantify, is a key notion for characterizing
 the relation existing between demographic dynamics and their possible resource limitation (7-9).
In here we will show that for almost one third of all world nations (i.e. water rich and VW dependent countries), the  carrying capacity depends on their food availability, which, in turn, depends on the available water resources. Therefore, a quantitative
 estimate  of the average local carrying capacity
$\bar{K}^i_{loc}$ of country $i$ is obtained  dividing the total
water currently available for food production in that country
(i.e., the current footprint of crops, grazing and livestocks
(20,21,26)), by the volume of water, $W^{i}_{c}$, used to produce
the food consumed on average by one individual in that nation
(16,19,20). $\bar{K}^i_{loc}$ is the maximum population
sustainable with the available local freshwater resources of
country $i$ and is expressed in terms of number of individuals.
The  virtual carrying capacity $\bar{K}^{j}_{V}$ is the maximum
sustainable population of country $j$ when  VW imports and exports
are accounted for. Thus, the virtual carrying capacity depends on
the structure of the global VW trade network (26,27). 
We also may incorporate in our modeling approach the impact on $\bar{K}$ of
changes in consumption patterns, crops expansion, and increase in
the efficiency of agricultural production (26).
\subsection{Relation between demographic growth and water avaliability}
We analyze and model the relation between demographic growth and
the water balance of water rich and VW dependent countries (for a
total of N=52 nations) using a stochastic logistic model. The
population dynamics, $x^i(t)$,  of nation $i$  are expressed as
(26)

\begin{equation}\label{logistic}
\frac{d x^i (t)}{dt}=\bar{\alpha}^i  x^i(t)\bigg(1-\frac{ x^i(t)}{K^i}\bigg),
\end{equation}

where $i=1,2,...N$. $K^i$ is treated as a random variable to
account for stochastic  fluctuations in available resources and
for uncertainties in their estimate. Thus, $K^i$ is expressed as
$K^i=\bar{K}^{i}+\xi$,  $\xi$ is a white Gaussian noise with mean
$\langle \xi \rangle=0$ and covariance $\langle \xi(t) \xi(s)
\rangle=\sigma^2_{K^i}\delta(t-s)$ (26).  $\langle \cdot \rangle$
denotes the average with respect to the stochastic fluctuations,
while $\sigma^2_{K^i}$ is the intensity of the fluctuations, and
$\delta(.)$ the Dirac's delta function. The model is based on two
key parameters: $\bar{\alpha}^i$ and $\bar{K}^{i}$. The mean
population growth coefficient $\bar{\alpha}^i$,
 is determined for  each nation using demographic data for the
period of record 1970-2010 (26-28). To assess whether the demographic
growth of a nation is driven by either local or virtual water
availability, we consider both the case of population dynamics
controlled by local water resources (i.e., with average "carrying
capacity", $\bar{K}^{i}=\bar{K}^{i}_{loc}$) and the effect of  VW trade
(i.e., $\bar{K}^{i}=\bar{K}^{i}_{V}$).

Figures 2-3 summarize  the results of this analysis. We stress
that the effective carrying capacities  are obtained
from direct evaluation of the actual virtual water availability and water footprint calculations (26) and not from fit of demographic data. The average
demographic growth of water-rich countries is well described by
the logistic model (Eq. (1)) with $K^i=\bar{K}^i_{loc}$, which
indicates that the growth of these populations is driven by local
water availability and is not limited by the important export of
VW sustained by these nations. Interestingly, several nations seem
to have just reached a critical phase in their demographic growth:
the point of separation between the local water and the VW regime
(see (26)). On the other hand, demographic growth of VW trade
dependent nations follows the logistic model with
$\bar{K}^i=\bar{K}^i_V$; thus, VW inputs may sustain the
demographic growth of populations that increasingly rely on
external VW resources (26). In these countries population growth
would not be sustainable without the import of VW from other
regions. The fact that demographic growth in water-rich countries
is independent of their VW exports, while water-poor nations
increasingly rely on VW imports highlights a situation that, in
the long run, will be unbalanced and unsustainable. These results
indicate that both water-rich and trade dependent populations are
growing relying on the same pool of resources. At some point it
will happen that, to meet their own demand, water-rich countries
will have to reduce their exports, thereby causing the emergence
of water limitations in trade dependent countries. Unless new
freshwater resources become available or investments for a more
water-efficient agriculture are made, these populations will have
to decrease.

\section{Sustainability of Future Scenarios}
To investigate  these effects we use the logistic model (Eq. (1))
coupled with the dynamics of food trade through the VW network
(Figure 3a). We first build a bipartite network with $N$ nodes
that interconnects $m$ water rich nations (blue nodes $r=1,2,..m$)
to VW trade dependent countries (red nodes $p=1,..,N-m$).
We investigate two scenarios. In the first one,the
population in blue nodes grows following Eq. (\ref{logistic}) with
$\bar{K}^r=(1-\beta)\bar{K}^r_{loc}$, where $\beta$ is the maximum
fraction of $\bar{K}^r_{loc}$ that water-rich countries are
willing to place  in the long term ($t \to \infty$) in the global
VW trade market.  At any time $t$, a water rich country, $r$, with
population $x^r(t)$  shares with trade dependent nations
a volume of water equivalent to $\bar{K}^r_{loc}-x^r(t)$; we
assume that this water is equally partitioned among all the $d_r$
trade dependent countries connected to that blue node (i.e., $d_r$
is the degree of node $r$). Therefore  the population $x^p(t)$ of
a red node, $p$, grows according to the Eq. (\ref{logistic}) with
virtual carrying capacity
$\bar{K}^p_V=\bar{K}^p_{loc}+\sum_{r=1}^m
a_{rp}(\bar{K}^r_{loc}-x^r(t))W^{r}_{c}/(W^{p}_{c} d_r)$, where
$a_{rp}$ is the adjacency matrix describing the virtual water
trade network ($a_{rp}$=1 if $r$ and $p$ are connected;
$a_{rp}$=0, otherwise). The case $\beta=0$ corresponds to the
purely competitive case, where in the long run $x^r(t) \rightarrow
\bar{K}^r_{loc}$ and no water is exported because all resources are
used to support the local population of the blue node $r$.
The second scenario takes into account plausible crop
expansions, increases in agricultural production efficiency, and
changes in diet and consumption rates (19,23,26), while the
cooperation regime is turned off ($\beta=0$). The relevant dynamics are
expressed by the same equations as in the previous case, but critically endowed with
time dependent local carrying capacities, $\bar{K}_{loc}(t)$, and
per capita virtual water consumption, $W_c(t)$. In particular, we
impose a growth rate in the local resources (i.e., in
 $\bar{K}_{loc}(t)$) to account for possible cropland expansions
and crop yield enhancement in existing croplands, and a decrease
in per capita virtual water consumption (i.e., in $W_c(t)$) due to
new practices that reduce water loss (19) and changes in consumption patterns (29).

\section{Discussion and Conclusions}
To evaluate the impact of the network structure, we investigate
the system's dynamics in the two above scenarios using
four  types of networks with different structural properties but
the same number of nodes and  average number of links. For
the first scenario, we find that in the purely competitive case
($\beta=0$)  the population of trade dependent countries increases
in the first 25 years  and decreases in the subsequent years as a
result of water limitations arising from the exclusion from the
access to water resources controlled and claimed by water-rich
countries (26,30). This decline continues in the following decades
as a results of demographic growth in water-rich populations
(Figure 3b). This behavior appears to be very robust and
independent of the topological properties of the underlying
network. If trade dependent countries use more VW resources than
they control, their population will strongly decrease once the
water rich countries start reclaiming all the water resources they
have access to, no matter how water is redistributed among the
trade dependent virtual water countries. The results are robust
and do not qualitatively change with different strengths of the
fluctuations of $K$ in the range $0 <\sigma_K<0.20 \bar{K}$.
Interestingly, this decrease in trade-dependent population is
reduced if a cooperative regime (i.e., $\beta>0$) is considered.
If water rich countries keep a fraction, $\beta>0$, of their water
resources in the VW global market, VW dependent countries can
sustain a larger population, which increases as a function of
$\beta$ (Figure 3b). The overall effect  of a cooperative regime
is a long-term increase in the total global population and
thus a  more sustainable demographic growth. In this cooperative
regime, the network topology affects the coupled VW-demographic
dynamics (26).  We finally note that the sensitivity of the
demographic dynamics on $\beta$ is very strong, and just a small
departure of $\beta$ from zero, may lead to substantial reductions
in the decrease of trade dependent populations (Figure 3b and
(26)).Also the second strategy, based on the enhancement
of productivity efficiency and a decrease in per capita global
consumption, results in a remarkable relief for trade dependent
countries, whose populations are subjected to less pronounced declines 
(Figure 3c). We also find that the increase in food (and virtual
water) availability resulting only through changes in consumption patterns and greater
equity in per capita consumption would not be sufficient anyhow  to meet the
increasing demand of a growing human population. In fact, in the
long run, if the growth rate of $\bar{K}_{loc}(t)$ tends to
zero, the trade dependent populations will peak and then
inevitably decline (26).

Despite the presence of a number of  other environmental,
cultural, and health related factors not included in this study,
this analysis points out how VW trade is only a temporary solution
to a local-to-regional unbalance between populations and food
production. The existence of this unbalanced condition might be
mitigated if a cooperative regime among water rich  and VW
dependent nations will continue to exist even once the excess of
VW in the exporting countries is strongly reduced by their
demographic growth ($\beta>0$). We finally  show that
strategies aiming at an increase in productivity efficiency
through agricultural practices that enhance crop yields while
reducing water losses (e.g., water harvesting, water conservation,
genetically modified crops) improve the sustainability of trade
dependent societies with respect to a decrease in export rates
from water rich countries.







\section*{References}
\begin{small}

1. Brown, L. R., Flavin, C., and Postel, S. (1991). Saving the Planet: How to Shape an Environmentally Sustainable Global Economy, The Worldwatch Environmental Alert Series.

2. Falkenmark, M. J., J. Rockstrom, and H. Savenjie
(2004).  Balancing Water for Humans and Nature, Earthscan, London.

3. Tilman, D., Balzer, C., Hill, J. et al. (2011). Global food demand and the sustainable intensification of agriculture.
PNAS, 108  (50), 20260-20264, doi: 10.1073/pnas.1116437108.

4. Barrett, C.B. (2010), Measuring Food Insecurity, Science 327 (5967), 825-828;
doi: 10.1126/science.1182768

5. Gebbers, R. and Adamchuk, V. (2010) Precision Agriculture and Food Security,
Science 327 (5967), 828-831; doi: 10.1126/science.1183899

6. Barnaby, W. (2009). Do nations go to war over water?
Nature, 458, 282. 283, doi:10.1038/458282a.

7. Ehrlich, P.R., Kareiva, P.M. and Daily G.C. (2012) Securing Natural Capital and
Expanding Equity to Rescale and Expanding Civilization. Nature 486, 68-73

8. Science for Sustainable Development (2012) Science 336 (1396)

9. Ehrlich, P.R. and Holdren, J.P. (1971) Impact of Population Growth, Science 171 (1212)

10. Rosegrant MW , M. Paisner, S. Meijer, J. Whitcover (2001) Global Food
Projections to 2020: Emerging Trends and Alternative Futures, IFPRI, Washington.

11. Rosegrant, M.W. and Cline, S.A.(2003). Global food security:
challenges and policies. Science 302 (5652), 1917-1919.

12. Arnell, N.W. (2004). Climate change and global water resources:
SRES emissions and socio-economic scenarios. Global Environmental
Change- Human Policy Dimensions, 14(1), 31-52.

13.Foley Jonathan A., et al. (2011). Solutions for a cultivated planet.
Nature 478 (7369), 337-342; doi: 10.1038/nature10452

14. Kumar, M.D. and Singh, O.P. (2005) Virtual water in global food and water policy making: Is there a need for rethinking? Water Resources Management, 19 (5), 759-789, 10.1007/s11269-005-3278-0.

15 Foley et al., (2005) Global Consequences of Land Use. Science 309 (570)

16.  Oki, T and Kanae, S. , (2004). Virtual water trade and world water resources.
Water Science and Technology, 47 (7), 03-209

17. Godfray HCJ, et al. (2010). Food security: The challenge of feeding 9 billion people.
Science 327 (5967) ,812�818;  2010 doi:10.1126/science.1185383.

18. Postel, S.L., Daily, G.C. and Ehrlich, P.R. (1996) Human Appropriation of Renewable Fresh Water. Science, 271 (785)

19. Rockstrom, J. Karlberg, J. and Falkenmark, M: (2011) Global
food production in a water-constrained world:exploring 'green',
'blue'challenges and solutions. Cambridge University Press,
Cambridge, U.K.

20. Hoekstra, A., and A. Chapagain (2008). Globalization of Water,
Wiley Blackwell, Malden, Mass.

21. Mekonnen, M.M. and Hoekstra, A.Y. (2012).  Water footprints of humanity.
PNAS 109 (9) 3232-3237, doi: 0.1073/pnas.1109936109

22. Allan, J. A. (1998). Virtual water: A strategic resource global
solutions to regional deficits. Ground Water, 36(4), 545.546,
doi:10.1111/j.1745- 6584.1998.tb02825.x.

23. Hanjra M.A. and M.E. Qureshi (2010). Global water crisis and
future food security in an era of climate change. Food Policy, 35,
365-377.

24. Rost, S., Gerten, D., Bondeau, A., Lucht, W., Rohwer, J.,
Schaphoff, S. (2008). Agricultural green and blue water consumption
and its influence on the global water system. Water Resources
Research. doi:10.1029/2007WR006331.

25. Fedoroff, N.V., Battisti, D.S., Beachy, R.N., Cooper, P.J.M.,
Fischhoff, D.A., Hodges, C.N., Knauf, V.C., Lobell, D., Mazur,
B.J., Molden, D., Reynolds, M.P., Ronald, P.C., Rosegrant, M.W.,
Sanchez, P.A., Vonshak, A., Zhu, J.-K., (2010).
Radically rethinking agriculture for the 21st century. Science 327 (5967)

26.  Materials and methods are available as supporting
material on PNAS Online.

27 Suweis, S, Konar M, Dalin C, Hanasaki N, Rinaldo A. and
Rodriguez-Iturbe I. (2011). Structure and controls of the global
virtual water trade network. Geophys. Res. Lett. 38 L10403. 8.

28. Wolfram, S.  (2011) Mathematica, version 8.0. CountryData Source Information.\\
http://reference.wolfram.com/mathematica/note/CountryDataSourceInformation.html

29. Arrow et al. (2004) Are We Consuming Too Much? Journal of Economic Perspectives, Volume 18, Number 3, 147-172

30. D'Odorico, P., F. Laio, and L. Ridolfi (2010). Does globalization
of water reduce societal resilience to drought?
Geophys. Res. Lett., 37, L13403, doi:10.1029/2010GL043167.

\end{small}


\begin{figure}
\begin{center}
\includegraphics[width=39pc]{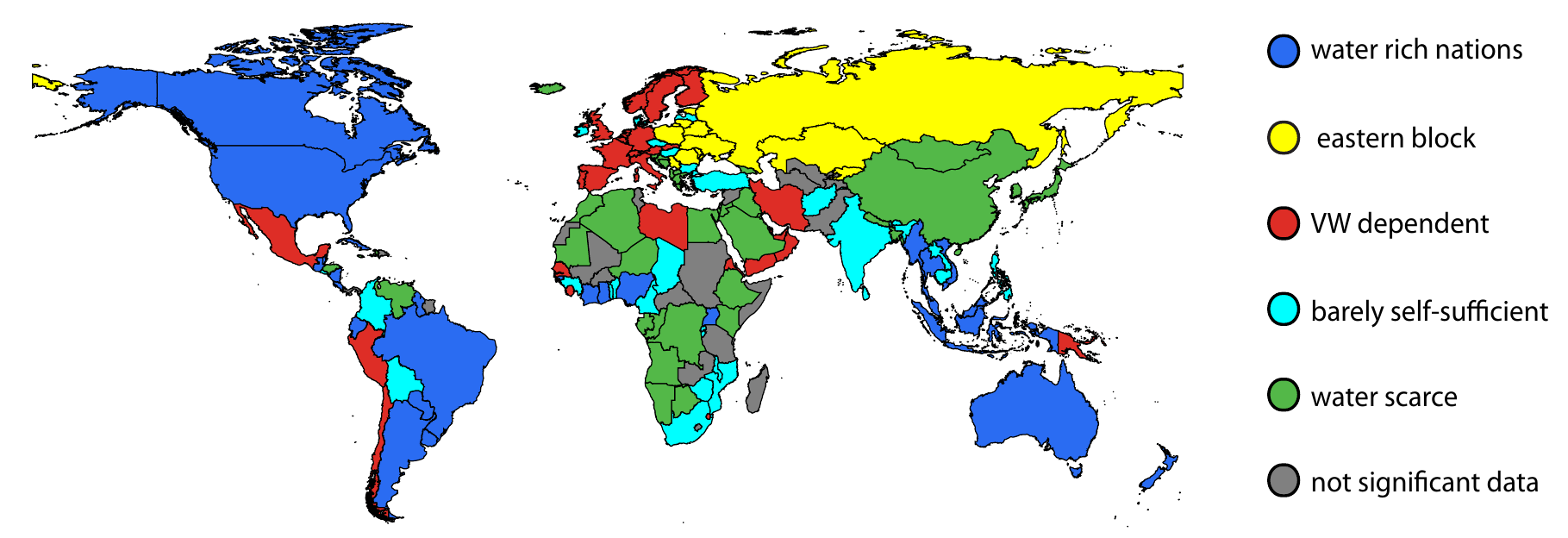}
\end{center}
\caption{Map of the World's nations classified on the basis of
their dependency on local and virtual water resources using data
for the 1996-2005 period (20,21). Countries are  {\it water rich}
when their mean population, $\bar{x}$, is less than  $0.8
\bar{K}_{loc}$; {\it virtual water dependent} if
$\bar{K}_{V} > \bar{x}>\bar{K}_{loc}$; {\it barely self sufficient} if
$\bar{K}_{loc}\approx \bar{x}$; {\it water scarce} if
$\bar{x}>\bar{K}_{V}>\bar{K}_{loc}$. Countries for which data do not appear to be consistent are shown in gray. A separate analysis has been carried out for the countries from the influence zone of the former Soviet Union (or the "eastern block") because in the past two decades their demographic dynamics have been affected by major political changes not related to water (26).}
\end{figure}

\clearpage

\begin{figure}
\begin{center}
\includegraphics[width=39pc]{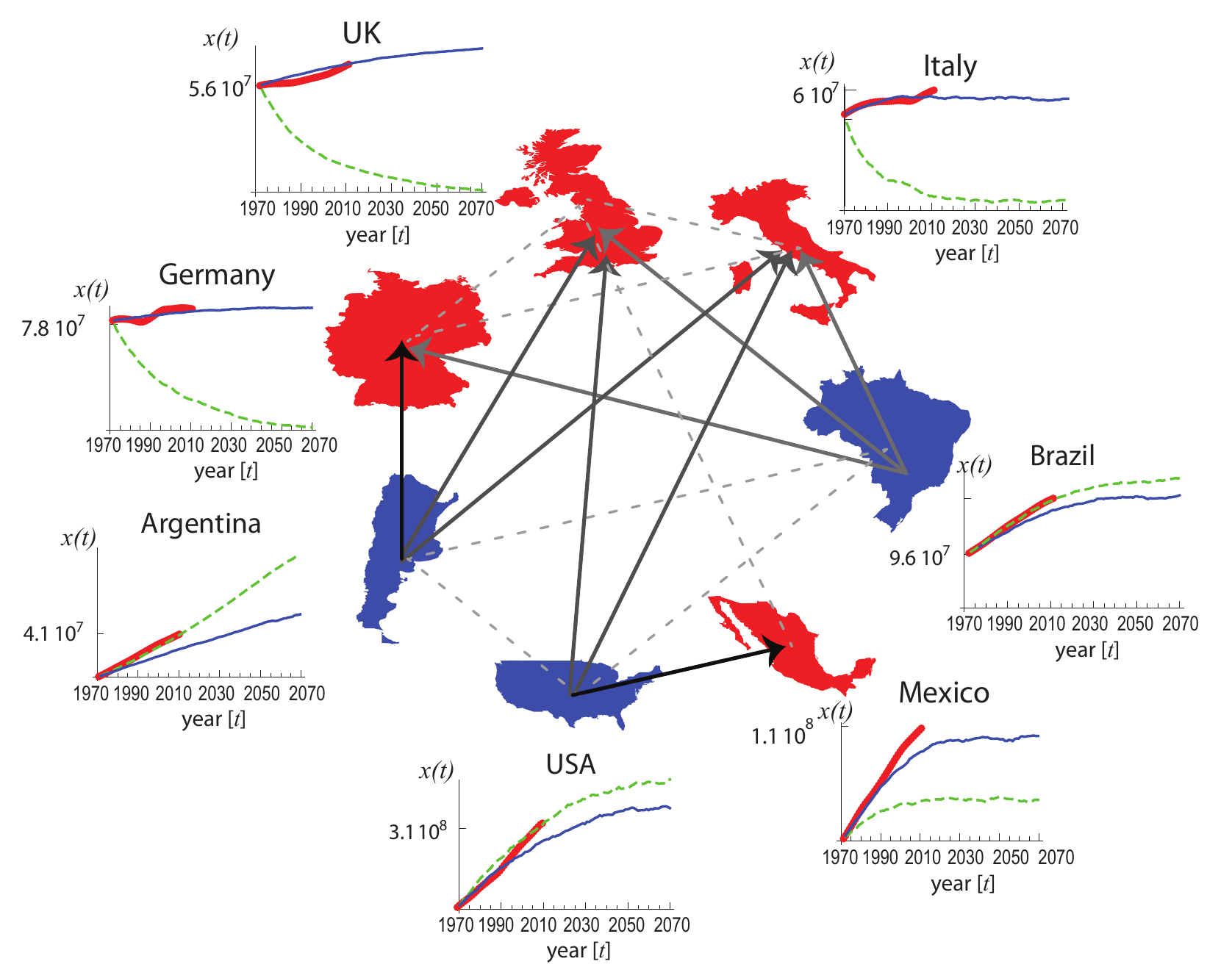}
\end{center}
\caption{Comparison between the logistic model given by Eq.
(1) and demographic data (red dots) from 1970 to 2011
(28) for three water-rich (blue) and four virtual water
dependent countries (red). These nations are connected
through the VW trade web.  The green dashed lines represent the
numerical simulations of Eq. (1) with carrying capacity $K^i=K^i_{loc}$, while the
blue lines corresponds to for virtual carrying capacity $K^i=K^i_V$.
Figures of the results for the complete analysis can be found in (26). The evolution of water rich countries population is predicted by Eq. (1) with $K^i=K^i_{loc}$, whereas for water dependent countries the appropriate evolution is given by Eq. (1) with $K^i=K^i_V$.}
\end{figure}

\clearpage
 
\begin{figure}
\begin{center}
\includegraphics[width=35pc]{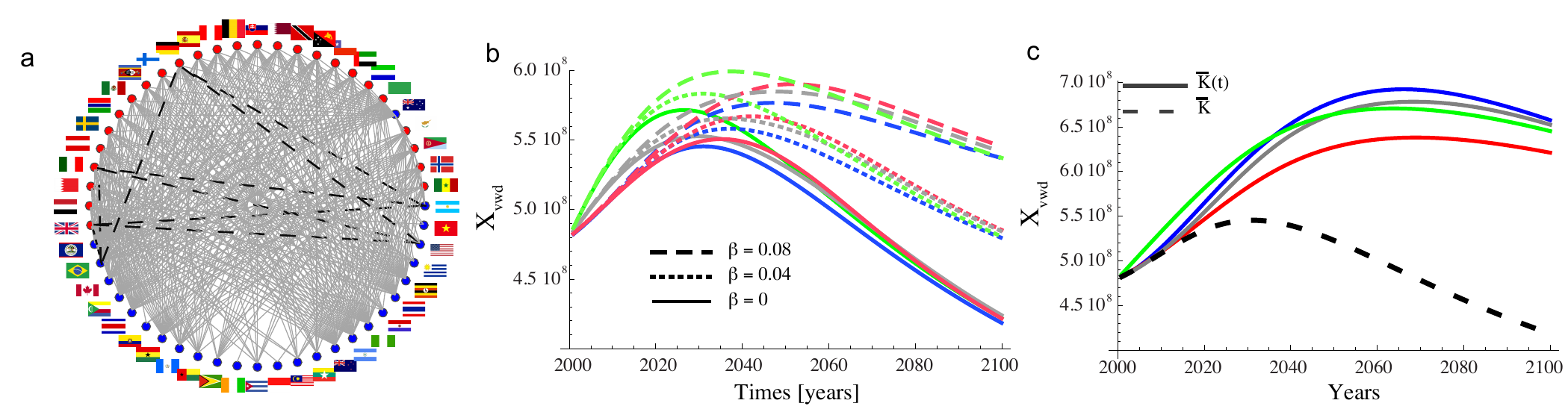}
\end{center}
\caption{a) Simulated coupled dynamical network. Blue nodes
represent water rich nations, while red nodes are the VW trade
dependent countries. We investigate the logistic model of coupled
population dynamics on four different kind of networks:
Random graph (26) (blue lines), small world network (26)
(red lines), scale free network (26) (gray lines)  and (in green)
the graph with topological properties similar to those observed in
the real global VW trade network (27) . b) Total population
$X_{vwd}$ of all VW dependent nations as a function of time in the
pure competitive case ($\beta=0$, solid lines), and in two
different cooperative regime ($\beta=0.04$-dot lines, and
$\beta=0.08$ dashed lines). c) Demographic
dynamics of VW dependent nations according to scenario 2 (solid
lines), which accounts for an increase in productivity efficiency,
crop expansion and decrease in consumption rates in the absence of
cooperation ($\beta=0$). The color scheme is the same as in panel
b. The dashed line corresponds to scenario 1 (panel b) with the
existing virtual water trade network (27) and with no
cooperation, $\beta=0$.}
\end{figure}






\end{document}